Type: Letter

# Evolution and spatial distribution of Brillouin backscattering associated to hybrid acoustic modes in sub-wavelength silica microfibers

*Wei Luo[1,2], Hong-qian Cao[2], Yong-kang Dong[3], Yan-qing Lu[2], Fei Xu[2],* and Gilberto Brambilla[4], **

*Corresponding Author: Fei Xu and Gilberto Brambilla*
E-mail: feixu@nju.edu.cn and gb2@orc.soton.ac.uk

[1]Center for Photonic Innovations, University of Electro-Communications, Chofu, Tokyo 182-8585, Japan
[2]National Laboratory of Solid State Microstructures, College of Engineering and Applied Sciences and Collaborative Innovation Center of Advanced Microstructures, Nanjing University, Nanjing 210093, China
[3]National Key Laboratory of Science and Technology on Tunable Lasers, Harbin Institute of Technology, Harbin 150001, China
[4]Optoelectronics Research Centre, University of Southampton, Southampton SO17 1BJ, UK

**Abstract:**

The spectral evolution and spatial distribution of backscattered Brillouin signals is experimentally investigated in sub-wavelength silica microfibers. The Brillouin spectrum evolution reveals the different dynamics of the various peaks, offering evidence of backscattering signals induced by acoustic waves with phase velocity greater than that of the longitudinal wave. The spatial distribution is found to have significant influence on the response of Brillouin scattering under tensile load, with hybrid acoustic modes providing a smaller response under axial strain. This insight into interactions between optical and hybrid acoustic modes at sub-wavelength confinements could help understand ultrasonic waves in tapered waveguides, and have potential applications in optical sensing and detection.

## 1. Introduction

Brillouin scattering (BS), which has been widely studied in the past decades[1-2], is a fundamental interaction between optical and acoustic waves, and has various applications ranging from spectroscopy[3] to distributed sensing of strain and temperature[4-5]. With the emergence and development of micro/nano-photonic devices, BS and other photon-phonon interactions have witnessed a renewed interest because of their unique characteristics displayed at the micro and nano scale. For instance, due to the tight confinement of both phonons and photons, hybrid acoustic waves (HAWs) arise in the small core of microstructured optical fibers, inducing a multi-peaked backward BS, which is quite different from what happens in standard single-mode fibers (SMFs)[6-8]. Moreover, strong photon-phonon coupling effects such as dynamic back-action[9-12], Brillouin cooling[13], on-chip BS[14] and tailorable stimulated Brillouin scattering (SBS)[15] have been reported in a variety of devices including optical microcavities, optomechanical crystals and nanoscale silicon waveguides.

Recently optical microfibers (MFs) have attracted much attention, as they provide sub-wavelength guiding and can be easily fabricated by a number of tapering techniques[16-17]. The low-loss confinement for both optical and acoustic waves, as well as the features of great configurability and easy connectivity to standard optical fibers, make this kind of tiny waveguide an ideal platform for observing opto-acoustic coupling effects. Unlike bulk acoustic waves, shear and longitudinal waves strongly couple in MFs, giving rise to a series of hybrid acoustic modes (HAMs)[18]. Acoustic waves propagating in these modes scatter the incident light, and each acoustic mode corresponds to a different Brillouin frequency shift for scattered light, which has been confirmed by previous experimental results of forward and backward BS in silica MFs[19-21]. The backscattering frequency shifts in MFs are generally lower than those in standard SMFs (~10.845 GHz at 1550 nm), and the minimum shifts is even smaller than 6 GHz for Rayleigh-like acoustic modes with large displacement near the surface of the fiber. Furthermore, BS can either be enhanced or cancelled through tailoring the photo-elastic and moving-boundary effects in MFs.

In this letter, we investigate the backward BS in silica MFs, presenting the backscattering spectrum evolution with diameter and the spatial distribution of Brillouin signals along a microfiber. Additionally, the response of backward BS under tensile load is also tested. Experimental results show previously unreported signals, as well as the complex strain response of BS influenced by the spatial distribution and acoustic modal properties, which may benefit both the fundamental understanding and potential applications for photon–phonon interactions at micro and nanoscale confinements.

## 2. Background

Brillouin scattering is initiated by broadband thermally-excited sound waves (acoustic phonons) which travel in acoustic modes in MFs. When propagating along the waveguide, these sound waves create moving index gratings which diffract the incident light meeting the phase-matching condition. The phase-matching condition for backward BS is $\beta_a = 2\beta$, where $\beta_a$ and $\beta$ are the sound wave and the optical wave propagation constants, respectively. According to this condition, Brillouin frequency shift for scattered light can be obtained from $v_B = 2n_{eff}V_a/\lambda$, where $n_{eff}$ is the optical mode effective refractive index, $V_a$ the acoustic phase velocity and $\lambda$ the optical wavelength in vacuum. Generally, all supported acoustic modes are able to satisfy the phase-matching condition, and backscattering by higher-order modes has a larger frequency shift. However, only the acoustic modes that have sufficient overlap with the optical mode can efficiently scatter the incident light[22]. In MFs, two acoustic mode families are usually considered to interact most efficiently with the fundamental optical mode, which are axially-symmetric radial ($R_{0m}$) modes and axially-asymmetric torsional-radial ($TR_{2m}$) modes, respectively. These two mode families are both HAMs because they involve both shear and longitudinal displacement components. Dispersion relation for $R_{0m}$ and $TR_{2m}$ modes are shown in Fig. 1 (a) and (b), and modal profiles of their fundamental modes are shown in the insets of the figures, respectively.

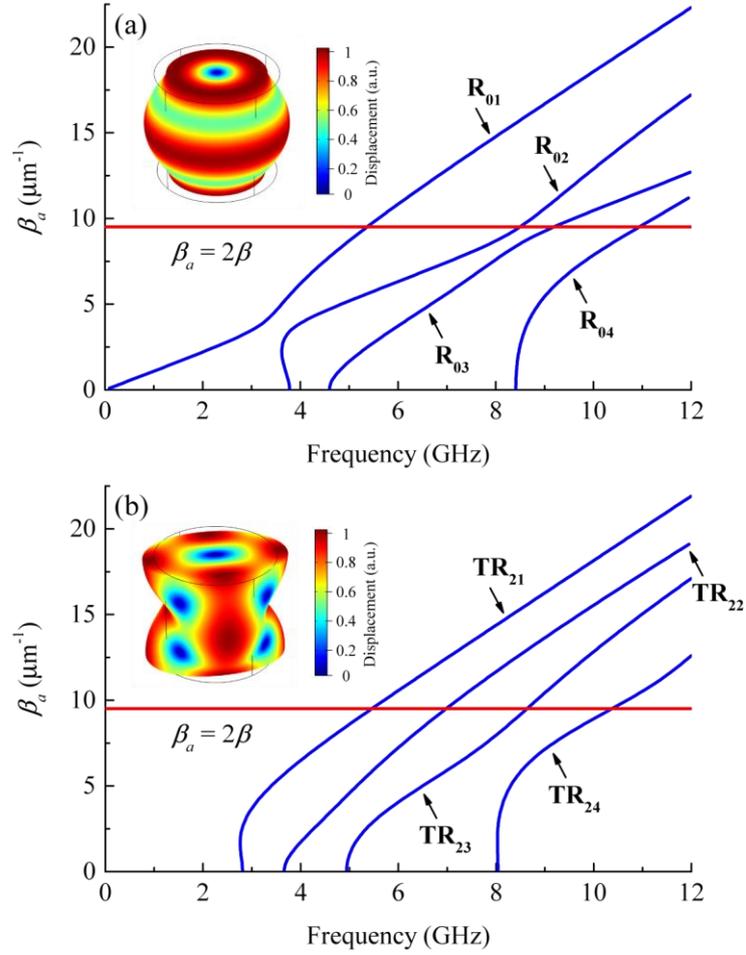

**Figure 1.** Characteristics of acoustic modes involving in backward BS. (a) Dispersion relation for $R_{0m}$ acoustic modes in a MF with 1-μm diameter. Horizontal red solid line represents the phase-matching condition for backward Brillouin scattering, while inset shows the displacement distribution of $R_{01}$ mode. (b) Dispersion relation for $TR_{2m}$ acoustic modes in a MF with 1-μm diameter. Inset shows the displacement distribution of $TR_{21}$ mode.

## 3. Experiments

### 3.1. Experimental setup

Figure 2 shows the schematic of the experimental setup used to measure the Brillouin backscattering spectrum. The output of continuous-wave (CW) tunable laser with linewidth of 100 kHz and wavelength set to 1550 nm is split into two beams by the

50:50 fiber coupler 1. One beam is amplified as pump wave, while the other beam serves as reference for heterodyne detection. The pump wave is then injected into the MF sample via an optical circulator, and generates backscattered signals. The reference and backward waves are then mixed through the 50:50 fiber coupler 2. One beam of the resulting mixed wave is coupled into a power meter for power monitoring, while the other beam gets received by an amplified photodiode (DC ~ 12 GHz). Finally, the Brillouin spectrum is recorded using an electrical spectrum analyzer.

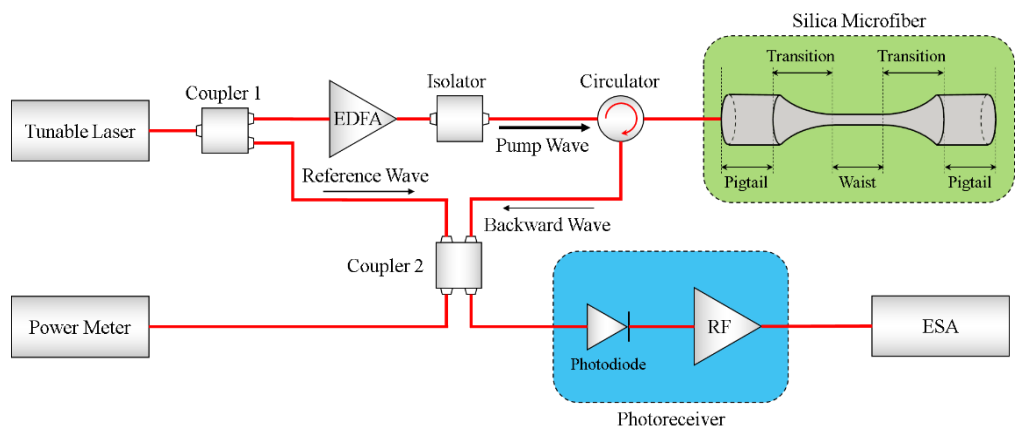

**Figure 2.** Experimental setup. The backscattered Brillouin signals from MF are detected using an optical heterodyne detection method. EDFA, erbium-doped fiber amplifier. RF, radio-frequency amplifier. ESA, electrical spectrum analyzer.

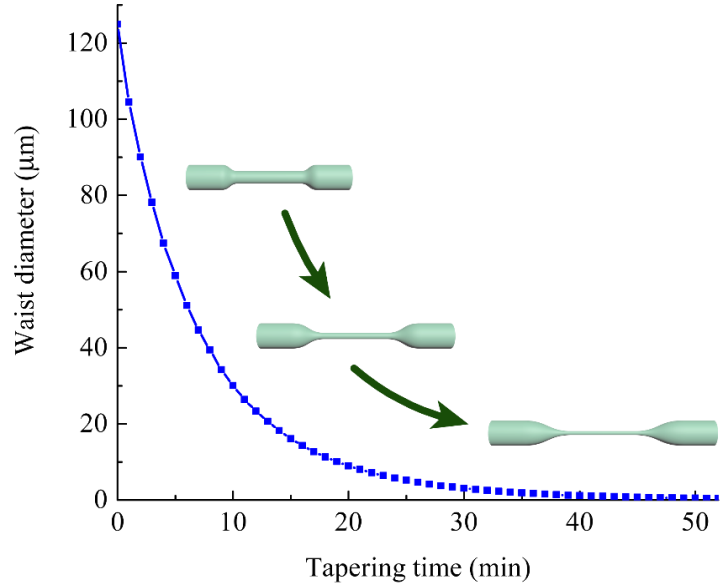

**Figure 3.** Relation between tapering time and waist diameter. Insets show that the shape of the fiber being tapered changes with the decrease of the waist diameter (not to scale).

**3.2. Sample preparation**

The schematic on green background in Fig. 2 shows the three-segment structure of the silica MFs investigated in the following experiments, which are tapered from standard SMFs (SMF-28) using the flame-brushing technique[17]. In the fabrication process, the fiber to be pulled is attached to two computer-controlled translation stages, and is softened in its central part by a fixed small hydrogen flame. The two translation stages keep moving back and forth at different rates, and elongate the fiber to create the MF with a shape determined by the trajectories of the two stages. Figure 3 shows the relation between tapering time and waist diameter during the fabrication process, which is based on the mass conservation law[23]. All the samples in our experiments were fabricated with a final waist length of ~19.5 mm and transition regions of ~64.9 mm on both sides.

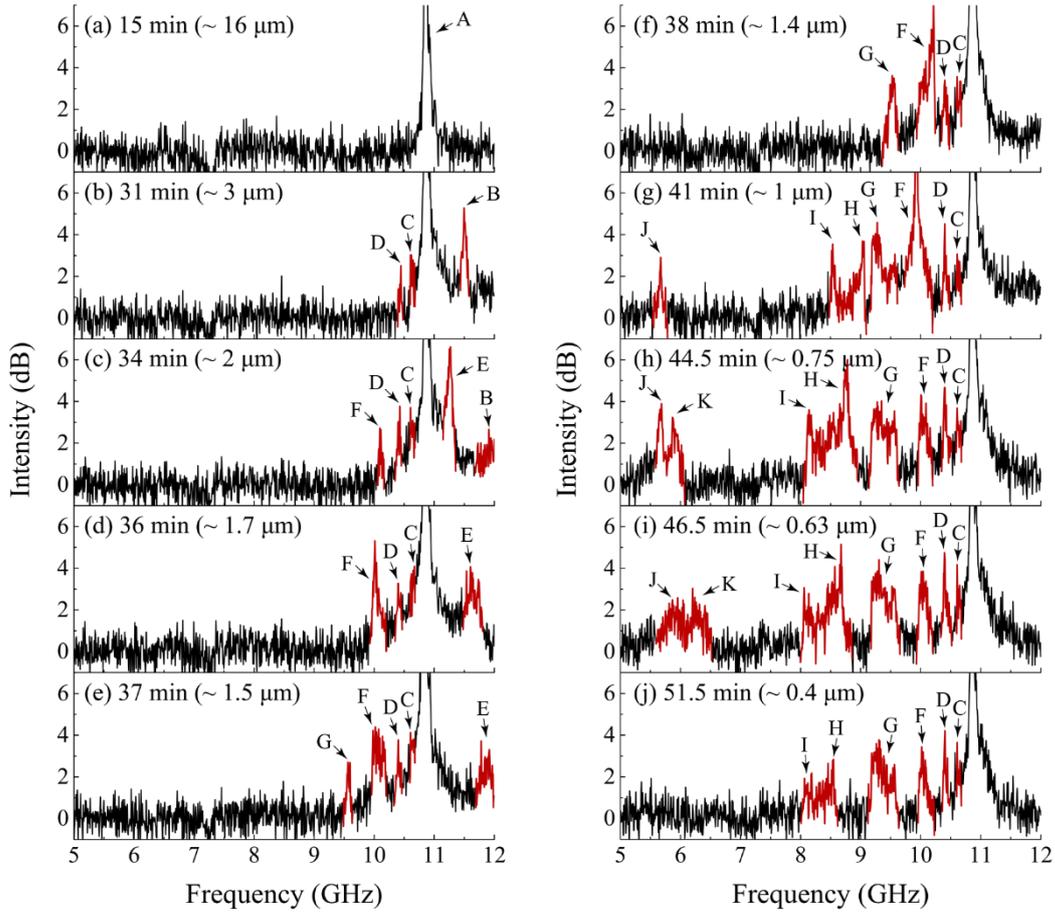

**Figure 4.** Brillouin backscattering spectra at (a) 15 min, (b) 31 min, (c) 34 min, (d) 36 min, (e) 37 min, (f) 38 min, (g) 41 min, (h) 44.5 min, (i) 46.5 min, and (j) 51.5 min during the fabrication of a MF sample. According to the relation in Figure 3, the corresponding waist diameters are estimated to be (a) 16 μm, (b) 3 μm, (c) 2 μm, (d) 1.7 μm, (e) 1.5 μm, (f) 1.4 μm, (g) 1 μm, (h) 0.75 μm, (i) 0.63 μm, and (j) 0.4 μm, respectively. Peak A generates from the SMF pigtails, and the red peaks labelled in rows from B to K are the signals that come from the tapered region. Signals above 12 GHz have not been measured due to the limited bandwidth of the photodiode.

### 3.3. Results and discussion

Backward Brillouin spectra were monitored in real time while fibers were tapered. Figure 4 shows the spectra at different times during the fabrication process of a MF sample. The corresponding waist diameter of the sample at each time node is also calculated and labelled in the figure, according to the relation shown in Fig. 3. In the beginning, the spectrum only shows a single peak (peak A) around 10.845 GHz, which represents the BS from the SMF pigtails. Afterwards, when the waist diameter is reduced to several microns, other scattering peaks start to arise at increasing lower frequency. The emergence of these peaks obeys the selection rule for acoustic modes. To be specific, the scattering strength is determined by the overlap between optical mode and acoustic modes, and the fiber diameter has a great impact on the modal profiles of both optical and acoustic modes. Therefore, the scattering strength highly depends on the diameter. Given that pump and backscattered waves propagate in the fundamental optical mode, higher-order acoustic modes reach a sufficient overlap with the fundamental optical mode at a larger diameter, and the corresponding scattering peaks emerge earlier in the spectrum. As the diameter further decreases, acoustic waves can no longer be well guided in the tapered fiber and the leakage dramatically increases, providing some wider peaks (peak B, E, J and K) that finally vanish at the end of the process. Peak C, D, F and G do not vanish because they finally originated in the transition region, which is the spatial distribution discussed below. Interestingly, BS peaks with frequencies higher than that of the SMF peak (i.e. B and E) have been observed, indicating scattering from high-order acoustic modes with phase velocities larger than that of a pure longitudinal acoustic wave (5970 m/s in silica). The shift towards higher frequencies increase for decreasing diameters. For example, the acoustic mode corresponding to peak B at 11.496-GHz in Fig. 4 (b) has a phase velocity of about 6364 m/s at a diameter of ~ 3 μm, while it migrates above 12GHz [Fig. 4(c)], with a phase velocity exceeding 6800 m/s at diameters smaller than 2 μm. This detail reveals one of the most important characteristics of HAMs in MFs.

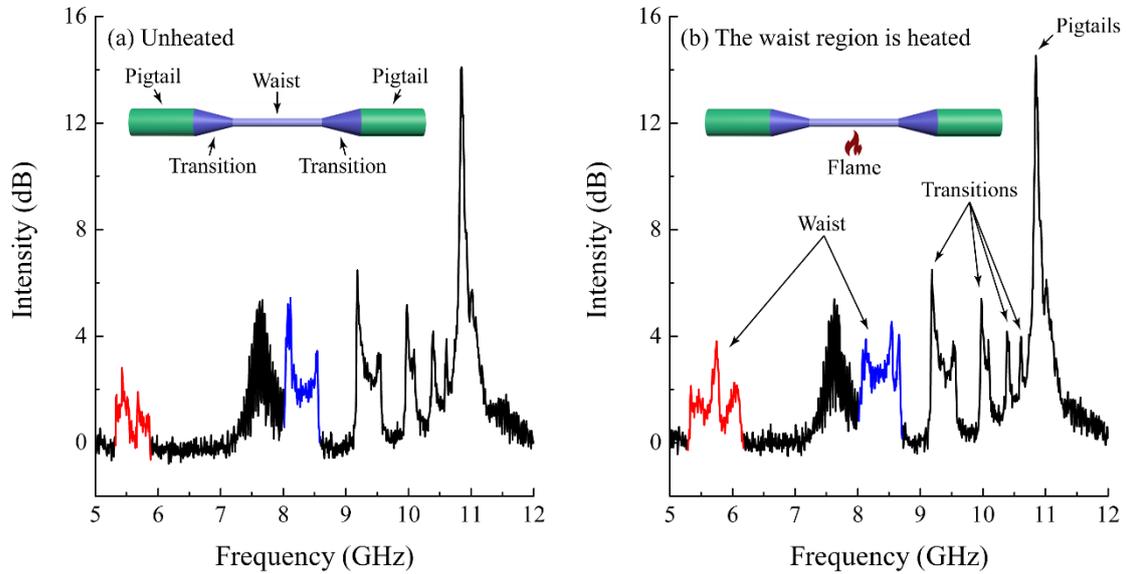

**Figure 5.** Heating test to determine the spatial distribution of the backscattered signals. (a) Brillouin spectrum of the sample when it is not heated and (b) when the waist region is heated. The MF diameter was ~0.76 μm.

Numerical simulations from previous work predicted the spectral distribution of the MF backscattered signals[21]. The evolution data provides information about the spatial origins of the signals. That is, in the last stage of the fabrication process [Fig. 4 (h) ~ (j)], the peak C, D, F and G at 9 ~ 10.8 GHz maintain their shapes without being affected by the diameter decrease. The only explanation for this phenomenon is that these signals are not generated from the waist region. A heating test was used to determine the BS spatial distribution, as the scattering peaks are sensitive to temperature. Figure 5 (a) shows the Brillouin spectrum measured in a MF sample with waist diameter of ~0.76 μm when the sample is not heated. The peak at 7 ~ 8 GHz is irrelevant to the BS and results from the noise of the laser source. According to Fig. 5 (b), when the waist region of the sample is heated, the peaks at 5 ~ 6 GHz (J and K) and at 8 ~ 9 GHz (H and I) experience changes in their shapes, indicating that these peaks owe their origins from the heated region. The same method was utilized to determine the origin of the peaks at 9 ~ 10.8 GHz (C, D, F and G) and showed that they originate from the transition regions.

If a tensile force is applied to a tapered fiber, due to the three-segment geometric structure, a non-uniform strain distribution is expected along the length. The deformation $\Delta L$ under a tensile force $F$ is calculated by

$$\Delta L = \int_0^L \frac{F}{G \cdot \pi \cdot r^2(l)} dl \tag{1}$$

where $L$ is the total length of the stretched region, $G$ is the Young's modulus of silica, and $r(l)$ is the fiber diameter along the length. According to Eq. (1), deformation of the waist, transitions and pigtails can be determined respectively. For samples with a waist length of ~19.5 mm, each transition length of ~64.9 mm and a distance between two anchor points of ~24 cm, an estimated 96.23% of the applied deformation is exerted on the waist, 3.75% on the transitions, and only 0.02% on the pigtails. Given the spatial distribution of BS in MFs, scattered signals located in the different regions should have diverse responses to this tensile force. In order to confirm this prediction, the response of BS under tensile load in a 0.73-μm MF was examined by stretching the sample with displacement steps of 0.1 mm, and the results are shown in Fig. 6. Figure 6 (a) shows that while the peak associated to the SMF fiber stands still, other peaks display different frequency shifts and linewidth variations with the same longitudinal deformation. The slopes of the linear fit for the frequency shifts of four major peaks are 425.7, 991.0, 611.0 and 347.1 MHz/mm, respectively. Peaks 1 and 2 have already been shown to originate from the uniform waist region. Therefore, it can be obtained that the actual response coefficients for peak 1 and 2 are 441.1 and 1026.9 MHz/mm, respectively. Considering the waist length of ~19.5 mm, the strain sensitivities of peak 1 and 2 are thus calculated to be 0.008609 and 0.02004 MHz/με, respectively.

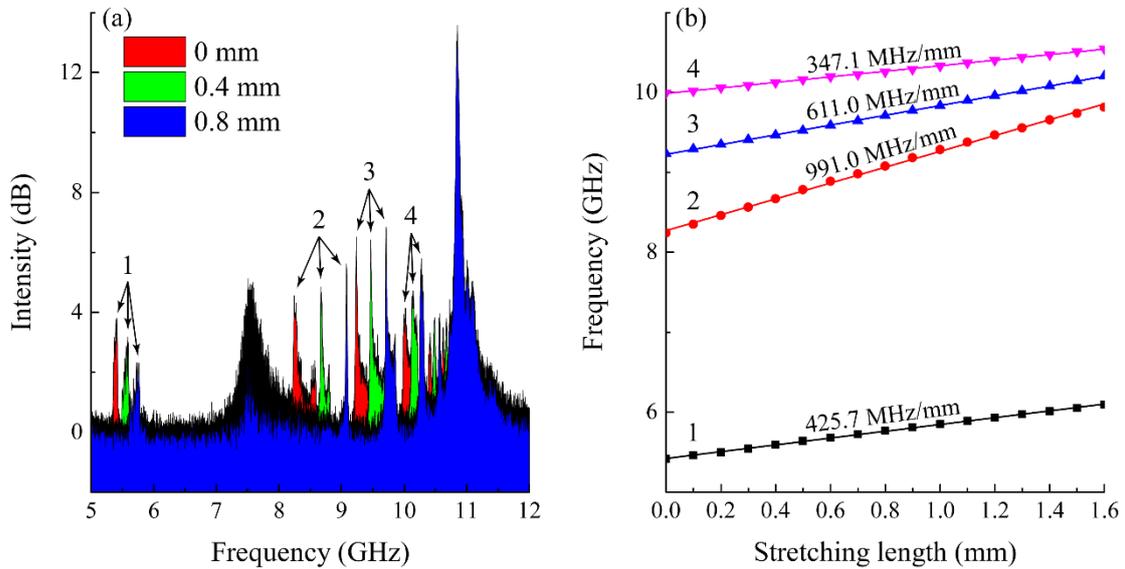

**Figure 6.** Response of Brillouin scattering under tensile load in a MF with ~0.73-μm waist diameter. (a) Brillouin spectra of the sample under three different displacements. (b) Frequency shift of four major peaks as a function of displacement. Black squares, red circles, blue up triangles and magenta down triangles respectively represent data for each peak, and the corresponding solid line of the same color stands for linear fitting.

The typical strain sensitivity for Brillouin scattering in SMFs is 0.0483 MHz/με [24], while peak 1 and 2 from the waist region exhibit 0.008609 and 0.02004 MHz/με, respectively. Acoustic modes in MFs show different strain responses from that of longitudinal acoustic modes in SMFs because of their hybrid nature. Starting from the theory of Brillouin scattering, the phase velocity of the acoustic wave corresponding to a certain scattering peak can be expressed as $V_a = v_B \lambda / 2n_{eff}$. For the investigated 0.73-μm MF without tensile strain, $\lambda = 1550$ nm and $n_{eff} = 1.0663$, thus $V_a(1) = 3910$ m/s and $V_a(2) = 5937$ m/s. From this estimation, $V_a(1)$ is close to the shear wave velocity for silica, while $V_a(2)$ approaches the longitudinal wave velocity. That clearly indicates that the shear wave component dominates in the HAM associated with peak 1, while the longitudinal wave component dominates in the HAM associated with peak 2. These results are consistent with our simulation (shown in Fig. 7), which demonstrated that the ratio of the shear displacement is 0.872 and 0.0385 for the $TR_{21}$ and $R_{02}$ modes associated with peak 1 and 2, respectively, and also seem to be in good agreement with the simulation results reported in previous work[20].

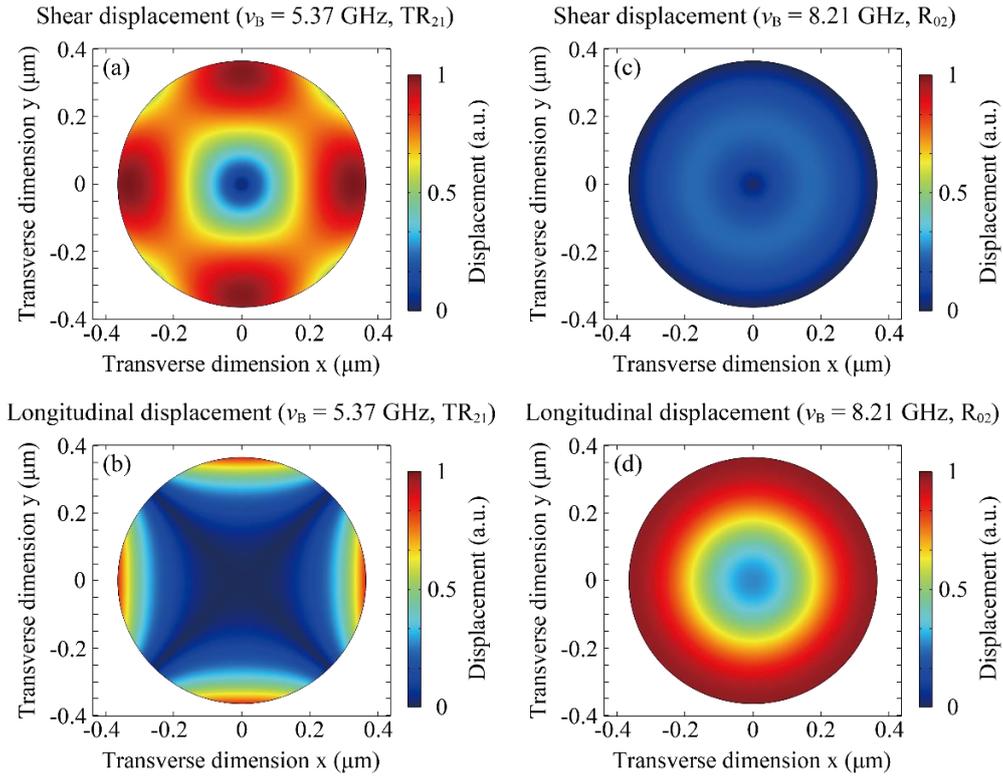

**Figure 7.** Shear and longitudinal displacements for the acoustic modes associated with peak 1 and 2 in the MF with 0.73-μm diameter. (a) Shear and (b) longitudinal displacements of $TR_{21}$ mode associated with peak 1 at 5.37 GHz. (c) Shear and (d) longitudinal displacements of $R_{02}$ mode associated with peak 2 at 8.21 GHz.

## 4. Conclusion

In conclusion, the spectral evolution and spatial distribution of backscattered Brillouin signals has been investigated in silica microfibers. The experimental results provide evidence of the selection rule for scattering from different acoustic modes, and verify the existence of Brillouin signals from HAWs with phase velocities beyond the compressional velocity 5970 m/s in silica. The peaks originated from the waist region have been discriminated from those originating in the transitions by a heating test. In addition, strain response of Brillouin scattering is examined in a stretched MF to explore the influence of the spatial distribution, and the diverse strain sensitivities of different acoustic modes has also been discovered. This work contributes to further comprehension of photon-phonon interactions in micro and nanoscale photonic devices, and the results may pave the way to potential applications for optical sensing and detection using BS of HAMs in MFs.

**Acknowledgements:**

This work was sponsored by National Key R&D Program of China (2017YFA0303700) and National Natural Science Foundation of China (61535005 and 61475069).

**References:**

1. Chiao, R. Y.; Townes, C. H.; Stoicheff, B. P., Stimulated Brillouin Scattering and Coherent Generation of Intense Hypersonic Waves. *Phys. Rev. Lett.* **1964,** *12* (21), 592-595.
2. Ippen, E. P.; Stolen, R. H., Stimulated Brillouin scattering in optical fibers. *Appl. Phys. Lett.* **1972,** *21* (11), 539-541.
3. Sonehara, T.; Tanaka, H., Forced Brillouin Spectroscopy Using Frequency-Tunable Continuous Wave Lasers. *Phys. Rev. Lett.* **1995,** *75* (23), 4234-4237.
4. Horiguchi, T.; Shimizu, K.; Kurashima, T.; Tateda, M.; Koyamada, Y., Development of a distributed sensing technique using Brillouin scattering. *J. Lightwave Technol.* **1995,** *13* (7), 1296-1302.
5. Bao, X.; Chen, L., Recent progress in Brillouin scattering based fiber sensors. *Sensors* **2011,** *11* (4), 4152-4187.
6. Dainese, P.; Russell, P. S. J.; Joly, N.; Knight, J. C.; Wiederhecker, G. S.; Fragnito, H. L.; Laude, V.; Khelif, A., Stimulated Brillouin scattering from multi-GHz-guided acoustic phonons in nanostructured photonic crystal fibres. *Nat. Phys.* **2006,** *2* (6), 388-392.
7. Beugnot, J.-C.; Sylvestre, T.; Maillotte, H.; Mélin, G.; Laude, V., Guided acoustic wave Brillouin scattering in photonic crystal fibers. *Opt. Lett.* **2007,** *32* (1), 17-19.

8. Kang, M. S.; Nazarkin, A.; Brenn, A.; Russell, P. S. J., Tightly trapped acoustic phonons in photonic crystal fibres as highly nonlinear artificial Raman oscillators. *Nat. Phys.* **2009,** *5* (4), 276-280.
9. Kippenberg, T. J.; Vahala, K. J., Cavity optomechanics: back-action at the mesoscale. *Science* **2008,** *321* (5893), 1172-1176.
10. Schliesser, A.; Del'Haye, P.; Nooshi, N.; Vahala, K. J.; Kippenberg, T. J., Radiation Pressure Cooling of a Micromechanical Oscillator Using Dynamical Backaction. *Phys. Rev. Lett.* **2006,** *97* (24), 243905.
11. Eichenfield, M.; Chan, J.; Camacho, R. M.; Vahala, K. J.; Painter, O., Optomechanical crystals. *Nature* **2009,** *462* (7269), 78-82.
12. Chan, J.; Alegre, T. P. M.; Safavi-Naeini, A. H.; Hill, J. T.; Krause, A.; Groblacher, S.; Aspelmeyer, M.; Painter, O., Laser cooling of a nanomechanical oscillator into its quantum ground state. *Nature* **2011,** *478* (7367), 89-92.
13. Bahl, G.; Tomes, M.; Marquardt, F.; Carmon, T., Observation of spontaneous Brillouin cooling. *Nat. Phys.* **2012,** *8* (3), 203-207.
14. Pant, R.; Poulton, C. G.; Choi, D.-Y.; McFarlane, H.; Hile, S.; Li, E.; Thevenaz, L.; Luther-Davies, B.; Madden, S. J.; Eggleton, B. J., On-chip stimulated Brillouin scattering. *Opt. Express* **2011,** *19* (9), 8285-8290.
15. Shin, H.; Qiu, W.; Jarecki, R.; Cox, J. A.; Olsson Iii, R. H.; Starbuck, A.; Wang, Z.; Rakich, P. T., Tailorable stimulated Brillouin scattering in nanoscale silicon waveguides. *Nat. Commun.* **2013,** *4*, 1944.
16. Tong, L.; Gattass, R. R.; Ashcom, J. B.; He, S.; Lou, J.; Shen, M.; Maxwell, I.; Mazur, E., Subwavelength-diameter silica wires for low-loss optical wave guiding. *Nature* **2003,** *426* (6968), 816-819.
17. Brambilla, G.; Finazzi, V.; Richardson, D., Ultra-low-loss optical fiber nanotapers. *Opt. Express* **2004,** *12* (10), 2258-2263.
18. Sriratanavaree, S.; Kejalakshmy, N.; Leung, D. M.; Rahman, B. M.; Grattan, K. T., Rigorous analysis of acoustic modes in low and high index contrast silica fibers. *Appl. Optics* **2015,** *54* (9), 2550-7.
19. Kang, M. S.; Brenn, A.; Wiederhecker, G. S.; Russell, P. S. J., Optical excitation and characterization of gigahertz acoustic resonances in optical fiber tapers. *Appl. Phys. Lett.* **2008,** *93* (13), 131110.
20. Beugnot, J. C.; Lebrun, S.; Pauliat, G.; Maillotte, H.; Laude, V.; Sylvestre, T., Brillouin light scattering from surface acoustic waves in a subwavelength-diameter optical fibre. *Nat. Commun.* **2014,** *5*, 5242.
21. Florez, O.; Jarschel, P. F.; Espinel, Y. A. V.; Cordeiro, C. M. B.; Mayer Alegre, T. P.; Wiederhecker, G. S.; Dainese, P., Brillouin scattering self-cancellation. *Nat. Commun.* **2016,** *7*, 11759.
22. Jen, C. K.; Oliveira, J. E. B.; Goto, N.; Abe, K., Role of guided acoustic wave properties in single-mode optical fibre design. *Electron. Lett.* **1988,** *24* (23), 1419-1420.
23. Birks, T. A.; Li, Y. W., The Shape of Fiber Tapers. *J. Lightwave Technol.* **1992,** *10* (4), 432-438.
24. Parker, T. R.; Farhadiroushan, M.; Handerek, V. A.; Roger, A. J., A fully distributed simultaneous strain and temperature sensor using spontaneous Brillouin backscatter.